\documentclass[lettersize,journal]{IEEEtran}

\usepackage{amsmath,amsfonts,amssymb,textcomp}

\usepackage{graphicx}
\usepackage[caption=false,font=normalsize]{subfig}

\usepackage{array}
\usepackage{booktabs}
\usepackage{tabularx}

\usepackage{algorithm}
\usepackage{algpseudocode}

\usepackage{url}
\usepackage{verbatim}
\usepackage{xcolor}
\usepackage[skip=0.5ex]{caption}
\usepackage{stfloats}
\usepackage{cite}
\usepackage[utf8]{inputenc} 
\usepackage[T1]{fontenc}    
\usepackage{amsmath}
\usepackage{amssymb}
\usepackage{multirow}
\usepackage{graphicx}
\usepackage{subfig} 
\begin{document}

\title{LecPrompt: A Prompt-based Approach for Logical Error Correction with CodeBERT}
\author{
    Zhenyu Xu\textsuperscript{\rm 1}, Victor S. Sheng\textsuperscript{\rm 1}\\
    \textsuperscript{\rm 1}Department of Computer Science, Texas Tech University\\
    \{zhenxu, victor.sheng\}@ttu.edu
}



\maketitle

\begin{abstract}
Logical errors in programming don't raise compiler alerts, making them hard to detect. These silent errors can disrupt a program's function or cause run-time issues. Their correction requires deep insight into the program's logic, highlighting the importance of automated detection and repair. In this paper, we introduce LecPrompt to localize and repair logical errors, an prompt-based approach that harnesses the capabilities of CodeBERT, a transformer-based large language model trained on code. First, LecPrompt leverages a large language model to calculate perplexity and log probability metrics, pinpointing logical errors at both token and line levels. Through statistical analysis, it identifies tokens and lines that deviate significantly from the expected patterns recognized by large language models, marking them as potential error sources. Second, by framing the logical error correction challenge as a Masked Language Modeling (MLM) task, LecPrompt employs CodeBERT to autoregressively repair the identified error tokens. Finally, the soft-prompt method provides a novel solution in low-cost scenarios, ensuring that the model can be fine-tuned to the specific nuances of the logical error correction task without incurring high computational costs. To evaluate LecPrompt's performance, we created a method to introduce logical errors into correct code and applying this on QuixBugs to produce the QuixBugs-LE dataset. Our evaluations on the QuixBugs-LE dataset for both Python and Java highlight the impressive capabilities of our method, LecPrompt. For Python, LecPrompt achieves a noteworthy 74.58\% top-1 token-level repair accuracy and 27.4\% program-level repair accuracy. In Java, LecPrompt delivers a 69.23\% top-1 token-level repair accuracy and 24.7\% full program-level repair accuracy. Especially when compared against other state-of-the-art models such as CoCoNuT, CodeBERTa, and RoBERTa, our method consistently showcases superior performance across multiple key metrics. These results illuminate the potential of LecPrompt in the domain of automated logical error correction in programming.

\end{abstract}


\section{Introduction}
logical errors, which are also known as semantic errors, can be a significant problem in programming. Unlike syntax errors, which produce compiler errors, logical errors occur when the code is written correctly but produces unintended or undesired results due to a mistake in the program's logic [1]. These errors can result from a programmer writing code that is not logily correct or when the program's execution flow is not as intended. Examples of logical errors in programming include using incorrect iteration numbers, using incorrect logical operators or comparisons, such as using "==" instead of "=", and working with incorrect data types, among other things. These errors can cause the program to function improperly, even if it is written correctly from a syntax perspective. For instance, using the wrong data type or an incorrect number of iterations can lead to unexpected results, making it difficult to identify and correct the problem [2].

Syntax errors are typically flagged by compilers or interpreters when the code is compiled or run. In contrast, logical errors are not detected by the compiler and can be more challenging to fix and localize. These errors often arise from incorrect assumptions or misunderstandings about how the program should behave. For instance,  a logical error is when a statement lacks a self-consistent logical structure after the condition. This can cause unexpected results and can make it difficult for the programmer to identify and correct the error. In such a case, the programmer must carefully analyze the logic of the program to determine the source of the error and how to fix it. Additionally, logical errors can manifest in unexpected ways, producing incorrect results for some inputs but not others. This makes it difficult for the programmer to identify the error and determine the most effective way to address it. To correct a logical error in programming, the programmer needs to carefully examine the code and identify the mistake. This often involves debugging the code, testing with different input values, or adding additional debugging statements. Once the error has been pinpointed, the programmer can fix the code and retest to verify that the issue has been resolved. However, this entire process can be very time-consuming and require a lot of effort.

It's important to note that repairing errors in a program does not always involve reasoning. Some errors can be easily detected and fixed by the compiler, such as syntax errors that violate the rules of the programming language. However, logical errors are more challenging to identify and repair. They occur when the code produces unintended or incorrect results due to a mistake in the program's logic, and require a more in-depth analysis and reasoning to fix. This may involve examining the code's algorithm, data, or other components to pinpoint the source of the error and make changes to the code accordingly. It's this process of programmatic understanding and reasoning that makes logical errors very difficult to fix.

Over the years, researchers and practitioners have proposed various approaches to tackle the problem of logical errors. One traditional solution is static analysis [3], which involves analyzing the program's code without actually executing it. This approach can be used to identify potential errors before the code is run, and can also be used to identify coding practices that are generally considered poor or unsafe. Static analysis tools can detect coding issues such as incorrect variable usage, incorrect output format, or data-type mismatches that can lead to logical errors. Another approach is dynamic analysis, which involves running the program with different inputs and observing its behavior to detect and diagnose errors [4]. This approach is useful in situations where the program is too complex for static analysis, or when the developer is unable to reproduce the error under normal circumstances. However, dynamic analysis can be time-consuming, and the testing process may not cover all possible input combinations. Program slicing is another approach that involves extracting a subset of the program's code that is relevant to a specific execution [5]. This approach can help narrow down the search for errors, making it easier to locate and fix logical errors. Besides the above-mentioned ways, there are several traditional approaches to diagnosing and fixing logical errors in software development, including model-based debugging, code review, code refactoring, and peer programming. Each of these approaches has its advantages and disadvantages, and selecting the most appropriate approach for a given project will depend on several factors, including the size and complexity of the program, the availability of resources, and the skills and expertise of the development team.



Recent advances in machine learning have shown promise in automatically detecting and correcting logical errors in code. Neural Machine Translation (NMT) models have been successfully applied to correcting syntax and logical errors by learning to translate incorrect code into correct code [6]. Supervised learning methods have also been used to classify code snippets as correct or incorrect, enabling developers to quickly identify and flag errors for correction [7]. Reinforcement learning techniques have also been employed to automatically generate code patches that not only fix the error but also maintain the program's correctness and functionality [8]. Large language models [9], such as GPT-3 [10] and BERT [11], have been shown to be effective for a wide range of natural language processing tasks, including those related to programming. These models are trained on large amounts of code data and can be fine-tuned on specific programming-related tasks, such as detecting and correcting logical errors. 

In this paper, we propose a prompt-based approach for automatic logical error correction (LecPrompt), an innovative method for token-level logical error correction. LecPrompt employs a prompt-based approach, tapping into the power of large language models to provide efficient and accurate logical error correction. Prompt learning adapts the downstream task of program logical error correction by fine-tuning inserted parameters while keeping the CodeBERT parameters frozen. We use soft prompts to transform program logical error correction tasks into MLM tasks, guiding the pre-trained model to effectively utilize programming language knowledge. In doing so, the model can adapt to the target language, maximizing the application of prior knowledge gained during the pre-training phase and significantly reducing training parameters. To assess LecPrompt, we also designed an iterative mask-filling approach that transforms QuixBugs into the QuixBugs-LE dataset, enriched with logical errors.
Our key contributions are:
\begin{itemize}
\item We are the first to employ a zero-shot localization method with LLM, using log probabilities to localize suspicious tokens and lines that may lead to logical errors.
\item We introduce an innovative iterative approach for creating Logical Error Datasets. Furthermore, we leverage CodeBERT's downstream task (MLM) to develop an automated logical error correction system compatible with Python and Java languages.
\item We present a soft-prompt solution for the high-cost logical error automatic correction task in a relatively low-cost scenario.
\end{itemize}

\section{Background}
\subsection{Fine-tune for Pre-trained Language Models}
Fine-tuning pre-trained language models involves adapting a model, which has already been trained on a large dataset, to a specific task or dataset. This adaptation allows the model to perform better on the target task by leveraging the knowledge it has acquired from its initial, extensive training. During this process, parameters of the model are adjusted based on the new dataset, optimizing its performance for the specific task.

\subsection{Prompt Fine-tune for Pre-trained Language Models}
Prompt fine-tune in natural language processing supplies additional information to a language model, helping it generate more precise outputs. This extra context, often used in in-context learning, enhances the model's understanding of tasks. Prompt fine-tune effectively harnesses pre-trained language models' capabilities for various tasks. For instance, when evaluating the sentiment of "Best pizza ever!", a template like "Best pizza ever! It was \_\_\_." transforms the sentiment analysis into a cloze task [22]. By creating suitable templates, pre-trained language models' potential can be effectively extracted. In addition to the hard prompt template approach, soft templates serve as an alternative method for fine-tuning processes, enabling the effective utilization of pre-learned knowledge from language models. As shown in Figure 1, we list the main idea of two different prompt methods:

\textit{Hard Prompt}: A hard prompt is essentially a discrete, predetermined input structure. Crafting these prompts mandates both domain expertise and an intimate understanding of both the language model and the targeted task. Due to their specific and manual design, they often excel in specific tasks for which they were intended but might lack the versatility required for broader applications or newly introduced tasks.

\textit{Soft Prompt}: Distinct from hard prompts, the soft prompt utilizes a continuous, tunable embedding structure. Rather than being manually designed, this vector is learned throughout the fine-tuning phase to adapt downstream tasks. A significant upside of this approach lies in its inherent flexibility and potential to discern intricate data patterns. However, a potential drawback is the resulting prompt might not provide the same interpretability as hard prompt.

\begin{figure}[ht]
\centerline{\includegraphics[scale=0.17]{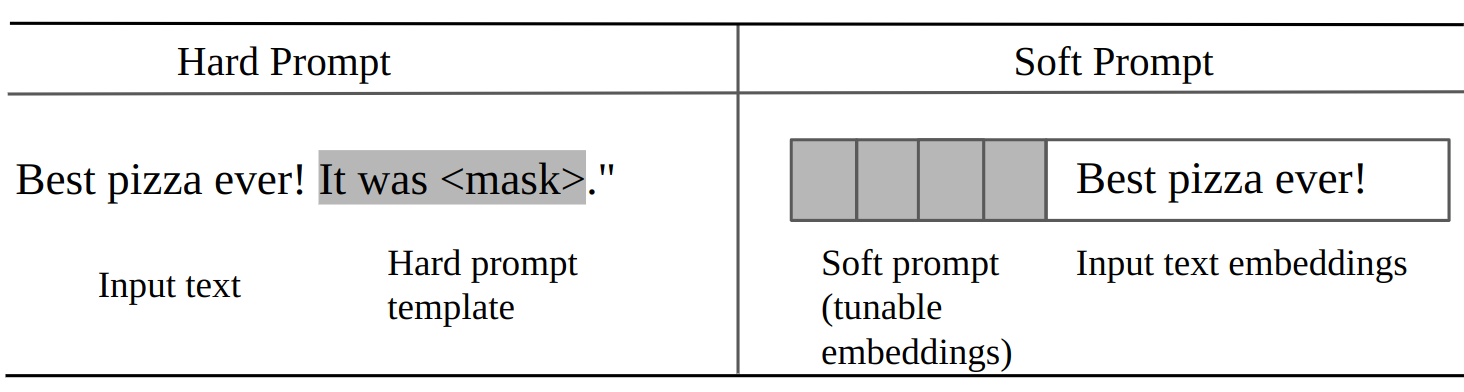}}
\caption{A comparison between hard and soft prompts. While hard prompts utilize a pre-designed template to elicit model outputs, soft prompts use continuously tunable embeddings to adapt various tasks. Where <mask>: delicious or disgust.}
\end{figure}

In this paper, we adapt soft prompt tuning as an alternative to traditional fine-tuning methods in language modeling for tackling the logical error correction task. Instead of designing specific templates, our approach involves inserting a prompt embedding layer into the input embeddings and fine-tuning only the added parameters, which is akin to prefix tuning. By leveraging large pre-trained models, we can fine-tune a limited number of prompt parameters to accomplish various downstream tasks without adjusting all parameters [26]. Consequently, soft prompt fine-tuning proves to be more space-efficient and requires lower hardware resources.


\section{Approach}
\subsection{Problem Description}
In the LecPrompt framework, the process is divided into two stages: localization and correction. In the localization stage, we use a log probability-driven approach to identify tokens in the program that likely contain logical errors. Once identified, these tokens are replaced with the placeholder <mask>. The correction stage employs CodeBERT. Having been trained on a vast dataset of program codes, CodeBERT can generate the appropriate token to replace each <mask>. However, it's designed to handle one <mask> at a time. To address multiple <mask> tokens, we use an autoregressive approach: after replacing one <mask>, the updated code is input back into CodeBERT for the next replacement. This process repeats until all placeholders are filled, producing the corrected program. To assess LecPrompt's effectiveness, we also develop a method that introduces logical errors into correct programs, simulating prevalent logical errors.

\subsection{Token-level Localization of logical error Programs}
To localize logical errors in code, we first need to understand two key concepts: perplexity and log probability. Perplexity (PPL) measures a model's uncertainty about a sequence, with higher values denoting more surprise or deviation from expected sequences. On the other hand, the log probability reflects a token's likelihood in its context, where values closer to 0 indicate a more common or expected token. LLMs tends to show a high perplexity about parts of the code where logical errors occur. The primary objective of our approach is to localize logical errors at the token level. Through a probabilistic lens, we aim to localize exact tokens in code sequences that are probable candidates for logical errors. This localization procedure has two layers of granularity: token-level and line-level.

For example, given a program below designed to compute the number of set bits in an integer. However, due to logical errors at two parts (i.e. \texttt{\^} and \texttt{-=}), it falls into infinite iterations:

\begin{center}
\begin{verbatim}
def bitcount(n):
    count = 0
    while n:
        n ^= n - 1
        count -= 1
    return count
\end{verbatim}
\end{center}

Figure 2 displays the log probabilities of individual tokens as calculated by language model. Tokens with significantly low log probabilities are potential indicators of logical errors. Figure 3 focuses on a line-level analysis, highlighting lines with high perplexities as likely error points. Notably, we exclude function definitions and package names, such as 'def function()' or 'import package', from our assessment, as they always have high perplexities but rarely contain errors. In our visualizations, blue bars represent tokens likely causing logical errors, while grey bars indicate overlooked tokens.

To systematically diagnose logical errors within code, we utilize the statistical might of mean and standard deviation applied to the log probabilities of code tokens. Starting with a given set of log probabilities, \(\{p_1, p_2, ..., p_n\}\), we calculate the mean, \(\mu\), and standard deviation, \(\sigma\), using the formulas:

\begin{align*}
\mu &= \frac{1}{n} \sum_{i=1}^{n} p_i \\
\sigma &= \sqrt{\frac{1}{n} \sum_{i=1}^{n} (p_i - \mu)^2}
\end{align*}

From here, we establish a threshold, \(\tau\), defined as \(\mu - k \times \sigma\). The factor \(k\) dictates the sensitivity of our anomaly detection, it's adjustable based on specific contexts. Tokens falling below this threshold are anomalies and potential sources of logical errors. This method's strength lies in its granularity: from analyzing individual tokens to entire code lines. Using this method on our data, we can effectively predict token and line indices potentially cause logical errors.

\begin{figure}[ht]
\centerline{\includegraphics[scale=0.45]{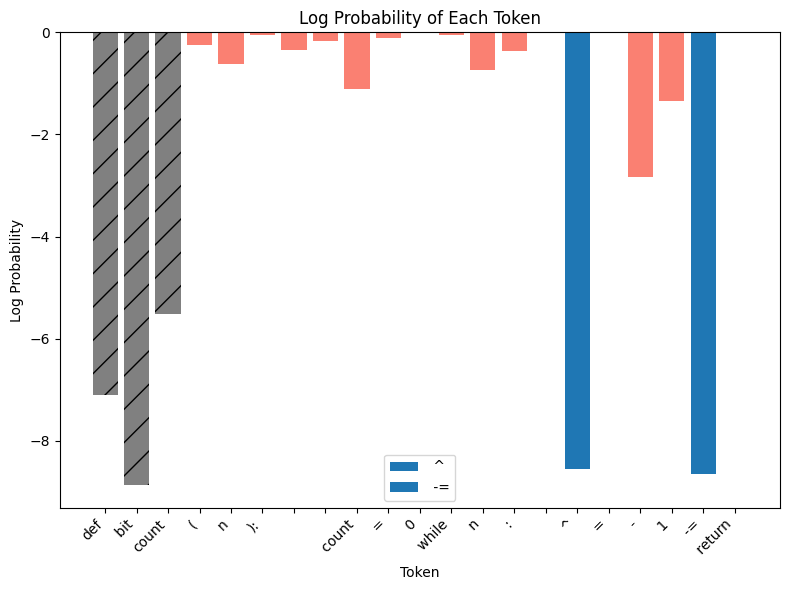}}
\caption{Log Probability of Each Token}
\centerline{\includegraphics[scale=0.45]{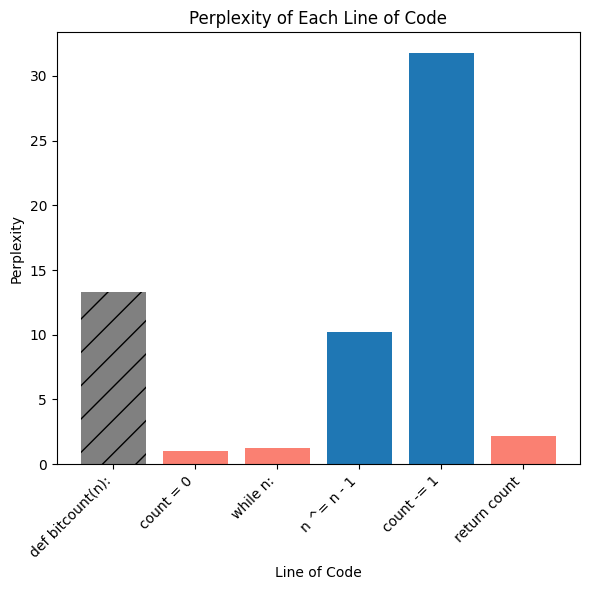}}
\caption{Perplexity of Each Line of Code. Blue bars highlight tokens potentially causing logical errors, whereas grey bars are intentionally disregarded tokens.}
\end{figure}

\subsection{Casting logical error correction to Masked Language Model (MLM) task}

After identifying the predicted logical error token, our subsequent objective is to rectify the masked tokens within the program. To cast the logical error correction task into a masked language modeling (MLM) task, we initiate the fine-tuning of the soft prompt embedding. Given an input text sequence $\tilde{x} = (x_1, x_2, ..., x_n)$, we employ a soft prompt embedding layer $\tilde{s} = (s_1, s_2, ..., s_m)$, where $m$ is the soft prompt length, and a text embedding layer $E_{\text{text}}(\tilde{x})$. These embeddings are then concatenated in the combined embedding layer 
$$E_{\text{comb}}(\tilde{x}, \tilde{s}) = [E_{\text{text}}(\tilde{x}); \tilde{s}].$$ 
The resulting combined embedding is passed to the frozen CodeBERT model 
$$h_1, ..., h_{|\tilde{x}, \tilde{s}|} = f_\theta ( E_{\text{comb}}(\tilde{x}, \tilde{s})),$$ 
which outputs contextualized representations of the input tokens, where $\theta$ denotes the CodeBERT parameterized by $\theta$, and its parameters are frozen. In the soft prompt finetune, our model only adapt parameters of soft prompt embedding. Subsequently, the hidden state associated with the input embeddings, denoted as $h_{-1}$, then is passed to a feedforward neural network (FFNN) with a Softmax activation function, which maps the hidden state to a probability distribution over the vocabulary. The output of the Softmax function is the predicted token for the MLM task: $$\hat{y} = \operatorname{softmax}(g_\phi (h_{-1}))$$

The MLM header $g_\phi$ is a neural network parameterized by $\phi$ that is optimized to minimize the cross-entropy loss: $$\mathcal{L}_{\text{MLM}}(\phi | y, \hat{y}) = -\sum_{i=1}^{|V|} y_i \log(\hat{y_i}),$$ where $y$ represents the true label for the MLM task, and $|V|$ signifies the size of the vocabulary.

Our subsequent step involves employing CodeBERT with soft prompt to predict all masked tokens in the program autoregressively. Each predicted masked token is filled into the original program and used as input for the next prediction. Each predicted masked token is filled into the original program and used as input for the next prediction. Figure 4 shows the process of using CodeBERT to predict and fill two <mask> tokens in an input program involves predicting the top candidate for each <mask> and replacing them accordingly, ultimately completing the correction process. 

\begin{figure*}[htbp]
\centerline{\includegraphics[scale=0.18]{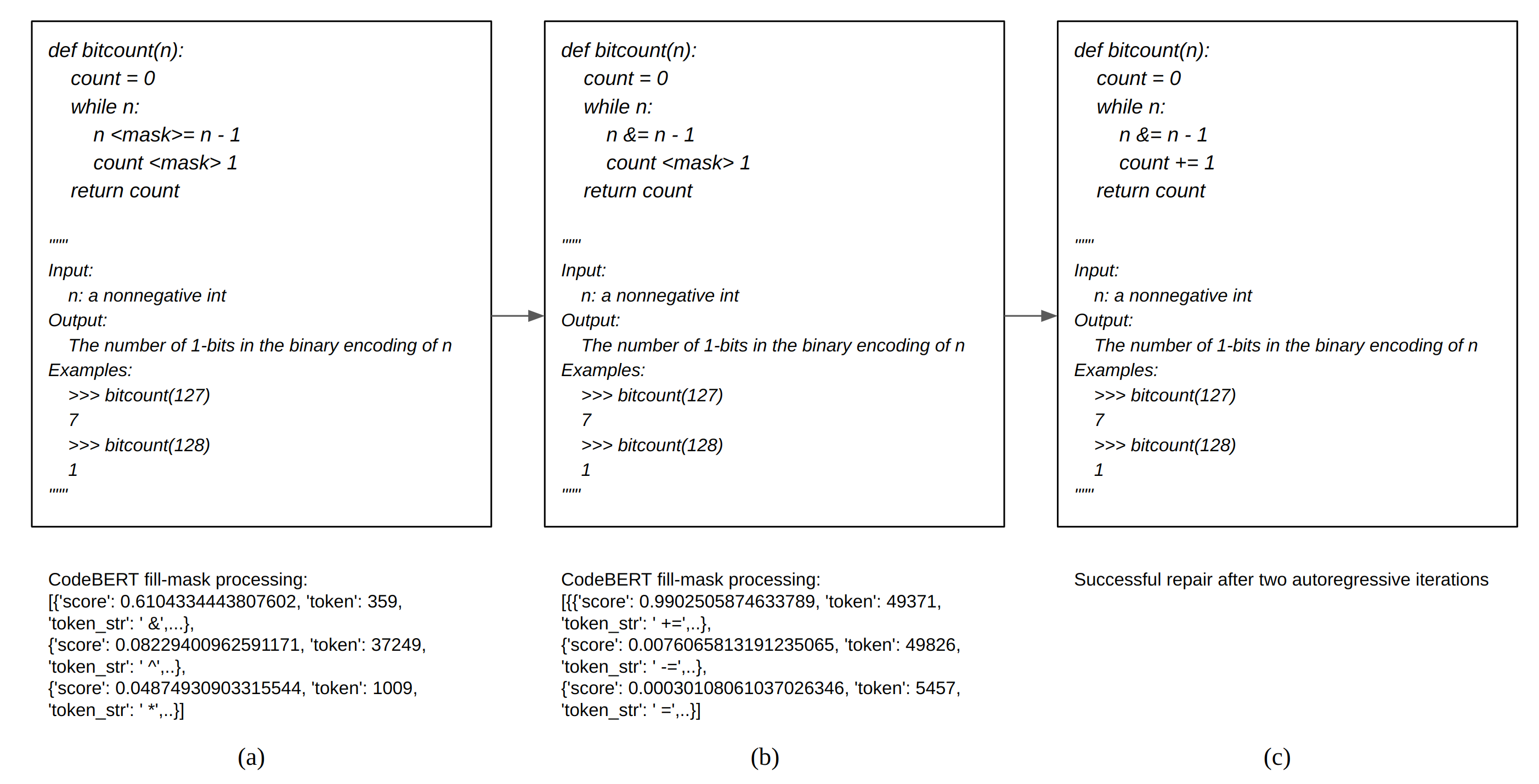}}
\caption{CodeBERT is used to predict and fill the two <mask> tokens in the input program. The process is as follows: (a) Use CodeBERT to predict the top 1 candidate for the first <mask> token, denoted as '\&', and substitute the first <mask> with '\&'. (b) Next, use CodeBERT to predict the second <mask> token, and substitute it with the top 1 candidate, denoted as '+='. (c) With both <mask> tokens now filled, the correction process is complete.}
\label{p1}
\end{figure*}

\subsection{Model Architecture}

Figure 5 illustrate the overall architecture of LecPrompt. The input to the model is a code snippet represented as text. This text is transformed into a word vector representation through an Embedding layer, which consists of two parts: a Tunable soft prompt and a Text embedding. The Tunable soft prompt is a learnable token sequence that is used to fine-tune the model for specific tasks. The soft prompt's weights are updated during training to better adapt to the task. The Text Embedding, on the other hand, transforms the input text into word vector representations. These representations form the original text embedding. The Tunable soft prompt and the Text Embedding are combined in the Combined Embedding layer to form a unified representation. This representation contains both the information from the input text and the task-related information provided by the soft prompt. The Combined Embedding is then passed as input to the CodeBERT model, where the parameters are frozen during fine-tuning. The CodeBERT model uses its Transformer [28] structure to learn the representation of the input code and complete the downstream task. Finally, the Combined Embedding, processed through the CodeBERT parameters, is passed to a feedforward neural network (FFNN) [29]. The FFNN transforms the embedding into prediction scores for each word in the vocabulary. The Softmax function is then applied to these scores to transform them into a probability distribution. The output is a probability distribution representing the model's prediction for each possible token. During prediction, the token with the highest probability is selected as the model's predicted output.

\begin{figure*}[ht]
\centerline{\includegraphics[scale=0.2]{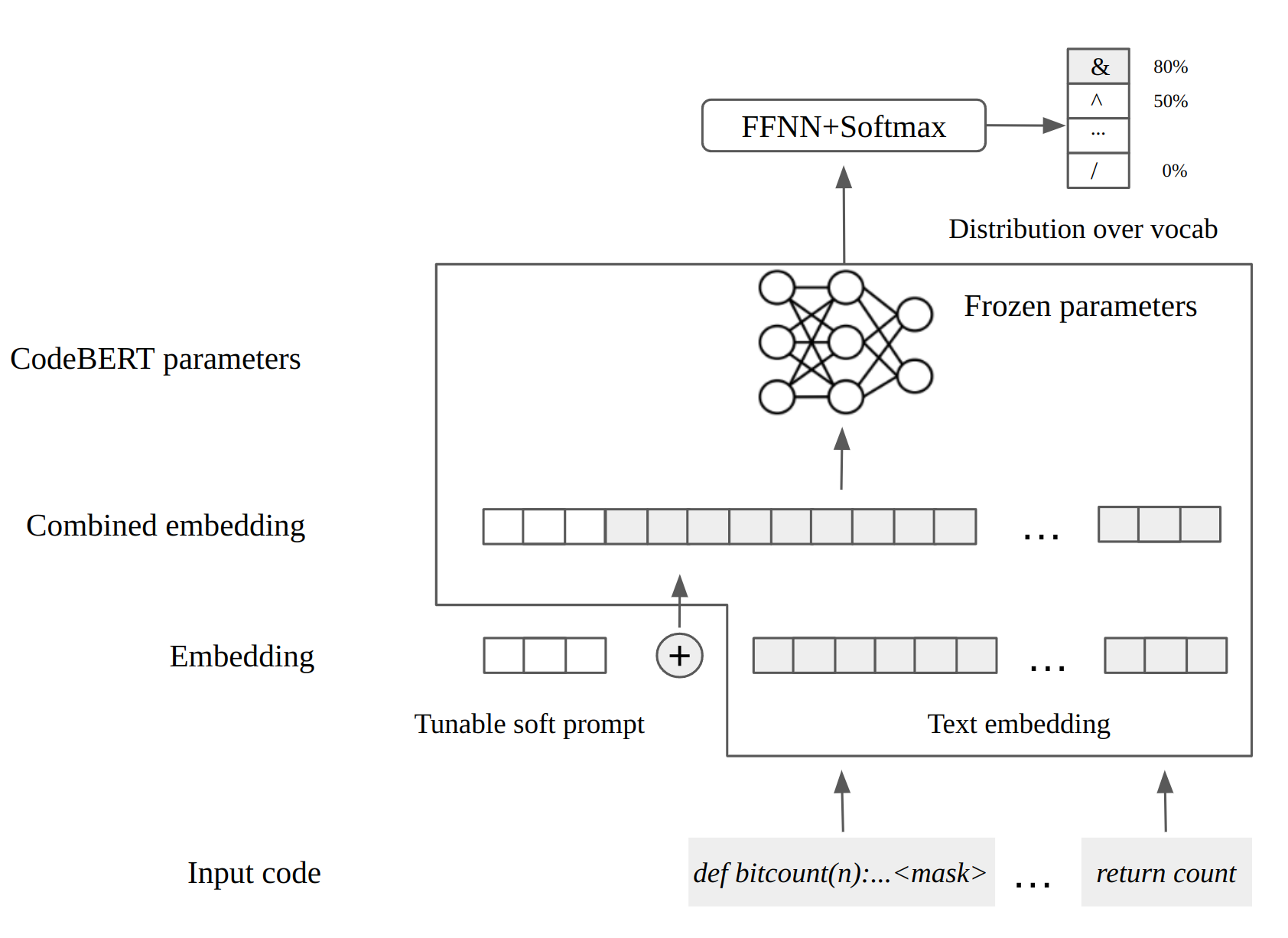}}
\caption{Model Achitecture}
\label{p1}
\end{figure*}

\subsection{Iterative Masking for Logical Error Introduction}
We introduce a novel iterative method for inducing logical errors into a codebase. Our approach integrates the concepts of replacing, deleting, and inserting random tokens with the `<mask>` token, drawing inspiration from the "fill-mask" procedures in prominent language models. The specifics of the algorithm are delineated in Algorithm~\ref{alg:logicalErrorIntro}. With a correct piece of code as the base, the method operates as follows:

\begin{enumerate}
    \item \textbf{Initialization}: Treat the initial code as the primary candidate.
    
    \item \textbf{Masking}: For each iteration, perform random operations: replace a token with `<mask>`, delete a token, or insert a `<mask>` token.
    
    \item \textbf{Filling the Mask}: Leveraging a predictive model, predict and replace each `<mask>` token, leading to an assortment of code variants.
    
    \item \textbf{Logical Error Checking}: Execute each variant and check for logical discrepancies. Only variants that run without syntax errors but produce incorrect results are retained.
    
    \item \textbf{Iteration}: The method iterates, intensifying the masking operations in every subsequent cycle, halting only when no new viable candidates that produce logical errors are found.
\end{enumerate}

\begin{algorithm}
\caption{Iterative Masking for Logical Error Introduction}
\label{alg:logicalErrorIntro}
\begin{algorithmic}[1]
\Procedure{IntroduceErrors}{code}
\State $\text{candidates} \gets [code]$
\State $\text{generation} \gets 0$
\While{variants produce incorrect results}
    \State $\text{newCandidates} \gets []$
    \For{each $\text{candidate}$ in $\text{candidates}$}
        \State $\text{masked} \gets \text{RandomMask}(\text{candidate})$
        \State $\text{filled} \gets \text{FillMask}(\text{masked})$
        \For{each $\text{variant}$ in $\text{filled}$}
            \If{not $\text{ThrowsError}(\text{variant})$ and produces incorrect result}
                \State append $\text{variant}$ to $\text{newCandidates}$
            \EndIf
        \EndFor
    \EndFor
    \State $\text{candidates} \gets \text{newCandidates}$
    \State $\text{generation} \gets \text{generation} + 1$
\EndWhile
\Return $\text{candidates}$
\EndProcedure
\Function{RandomMask}{code} \Comment{Returns code with random tokens masked}
\State \Return $\text{maskedCode}$
\EndFunction
\Function{FillMask}{maskedCode} \Comment{Returns a list of filled code variants}
\State \Return $\text{filledVariants}$
\EndFunction
\Function{ThrowsError}{code} \Comment{Returns True if the code throws an error, otherwise False}
\State \Return $\text{errorStatus}$
\EndFunction
\end{algorithmic}
\end{algorithm}

In this process, we prefer the lower score candidates for <mask>. Central to the efficacy of this method is its iterative nature that incrementally introduces complexity, ensuring the resultant code variants are syntactically correct. Moreover, the iterative approach not only produces relatively complex multiple logical errors but also simulates operations like insertion, deletion, and replacement.

\section{Dataset and Evaluation}
\subsection{Research Questions}
We aim to answer following research questions through experiments:

\textbf{RQ1:} How effective is LecPrompt in localizing and repairing logical errors compared to other state-of-the-art method?

\textbf{RQ2:} How does the design of the soft prompt and the mask percentage affect LecPrompt's performance?

\textbf{RQ3:} What advantages does the soft prompt method offer, and what are its limitations?

\textbf{RQ4:} What implications arise if the localization process misidentifies the location of a logical error?

We designed RQ1 to evaluate the performance of LecPrompt against various repair methods using the QuixBugs-LE dataset. RQ2 investigates the influence of the soft prompt's initialization, placement, and length, as well as varying mask percentages on LecPrompt's performance. Through RQ3, we aim to highlight the space-efficiency and few-shot scenario of the soft prompt approach and delve into its potential constraints. Lastly, RQ4 focuses on analyzing the potential ramifications of incorrect identifications during the logical error localization step.

\subsection{Implementation Details}
\subsubsection{Dataset Description}
To train and evaluate LecPrompt, we use two datasets: Codeparrot and QuixBugs-LE.

\textbf{Codeparrot}: Codeparrot, sourced from GitHub [31], is a vast collection of code spanning multiple programming languages. For LecPrompt's training, we employed a specific subset of Codeparrot: the GitHub-code-clean subset. We further refined the dataset, aiming for an average line length below 60, a maximum line length under 100, and ensuring that the fraction of alphanumeric characters exceeded 0.45. The resulting Codeparrot-filtered subset contains about 745,000 Python and 884,000 Java files. During the fine-tuning process, the CodeBERT model is trained on the Codeparrot-filtered subset using the masked language modeling task, where certain tokens are masked and the model is tasked with predicting the missing token.

\textbf{QuixBugs-LE}: The LecPrompt utilizes the QuixBugs-LE dataset to assess its performance. The original QuixBugs dataset, sourced from the Quixey Challenge, comprises 50 Python and 40 Java programs [30]. For our dataset modification, we introduced logical errors into the correct Python and Java programs using an iterative masking method. A comprehensive description of this method is available in Section 3.B. Though these modified programs run as expected, they fail the provided test cases. We set a chunk size of 20, ensuring logical errors disperse across various lines, deliberately excluding the program's beginning and end due to their critical role for model comprehension.  Furthermore, we remove irrelevant comments, keep only the task-relevant code. This ensures that CodeBERT, which has been trained on both natural and programming language inputs, can utilize comment text during code repair. After modifications, the QuixBugs dataset evolved into the QuixBugs-LE dataset, hosting 550 Python and 662 Java programs. On average, each program contains 2.6 logical errors.

\subsubsection{Localization}
Basing on our strategy, we used \textit{text-davinci-003} to calculate perplexity (PPL) and log probability and localize logical errors. A sensitivity factor \( k \) was set at 1.5.

\subsubsection{Repair}
We employ the default tokenizer of CodeBERT (i.e., Microsoft/codebert-base-mlm) with a vocabulary size of 50,265. We configure the maximum sequence length to be 512. When dealing with programs exceeding this length, we utilize a sliding window technique to ensure that CodeBERT effectively captures the semantic information surrounding the masked token. The percentage of the training dataset to be used as the validation dataset to 5\% when there is no separate validation set provided. Our models are built upon the widely-used CodeBERT (H=768, A=12, L=12) as our base model. H represents the number of self-attention heads, A represents the number of layers, and L represents the dimension of the hidden state. During the pre-training phase, CodeBERT learns to represent various programming languages, including Java, Python, JavaScript, PHP, Ruby, and Go. We continued to finetune it with the masked language modeling (MLM) task on Python corpus. 
 
Our models are optimized using the AdamW [32] algorithm and trained with a batch size of 16.  We employ a batch size of 16 and an initial learning rate of 5e-5, which linearly decays to 0 throughout the training process. We employ soft prompt fine-tuning and combine the soft prompt embedding uniformly with the original CodeBERT embedding. During training, we periodically evaluate model performance on the validation set and select the checkpoint with the best accuracy for testing. We implement our experiments using the Transformers library and runs on a machine with two GeForce RTX 3080 GPUs.
\subsubsection{Baselines}
We compare our approach with following baseline models.

\textbf{CoCoNuT:} CoCoNuT is a new automated program repair technique that uses ensemble learning and context-aware neural machine translation [40]. It uses convolutional neural networks (CNNs) to extract hierarchical features, better modeling source code at different granularity levels, and takes advantage of the randomness in hyperparameter tuning to build multiple models, using ensemble learning to fix more errors. CoCoNuT has been successfully applied to Quixbugs dataset, and shows its potential to repair logical errors.

\textbf{RoBERTa}: RoBERTa [33] is a large-scale pre-trained language model that utilizes the transformer architecture. It was created by Facebook AI Research as an upgrade to the original BERT (Bidirectional Encoder Representations from Transformers) model. RoBERTa shares the same foundational architecture as BERT, but distinguishes itself through several significant modifications to its training process, leading to enhanced performance across a broad spectrum of natural language processing tasks.


\textbf{CodeBERTa}: CodeBERTa [35] is a language model that has been designed specifically for processing source code. It is a variant of RoBERTa that has been fine-tuned on a large corpus of code from Github and StackOverflow, which enables it to learn the unique patterns and relationships inherent in programming language syntax and structure. This specialization makes CodeBERTa highly effective for tasks such as code completion, code generation, and code analysis.

\textbf{CodeBERT}: CodeBERT [12] is a state-of-the-art pre-trained language model that has been specifically designed for source code processing. It is based on the Transformer architecture and has been fine-tuned on a large corpus of code from Github and StackOverflow.

\subsubsection{Fine-tuning and Soft prompt-tuning}
We specifucally use microsoft/codebert-base-mlm as our base model. CodeBERT-base-mlm is pre-trained using only the masked language modeling (MLM) task. In the MLM task, certain tokens in the input sequence are randomly masked and the model is trained to predict the missing tokens. We continue to fine-tune on CodeBERT models with following two ways:

CodeBERT-MLM-\textbf{FT}: CodeBERT-MLM-\textbf{FT} involves fine-tuning the models for 800,000 steps using a batch size of 16 on python programs from the Codeparrot-filtered dataset. This fine-tuning is performed on the masked-language-modeling task, where 5\% of the Codeparrot-filtered dataset is reserved for use as the validation dataset. During the CodeBERT-MLM-\textbf{FT} process, we fine-tune the parameters of the pre-trained CodeBERT model to better suit our MLM task. Unlike in the soft prompt approach, we do not use any tunable soft prompt during the CodeBERT-MLM-\textbf{FT} process.

CodeBERT-MLM-\textbf{PT}: CodeBERT-MLM-\textbf{PT} is a fine-tuned model where we only update the weights of the trainable soft-prompt embedding, while the CodeBERT parameters are frozen. During the training process, we set the prompt length to 40 tokens and use a mask percentage of 55\% for the masked-language-modeling task. The training dataset consists of Python programs from the Codeparrot-filtered dataset, and we use 5\% of this dataset as the validation dataset. We will discuss the choices of prompt length and mask percentage in the Discuss section.

\subsection{Metrics}

To comprehensively assess our approach's effectiveness, we have established a set of metrics for both error localization and repair.

\subsubsection{Localization Evaluation}
At the token-level, we gauge the accuracy with which our model identifies the specific tokens within code lines that contain logical errors. In contrast, at the line-level, we evaluate the model's precision in localizing which lines of code contain logical errors, without find out the specifics of which token within that line is erroneous.

\subsubsection{Repair Evaluation}
Given a exact logical error token locations, and we have the token-level repair accuracy for the top 1/5/10 repair candidates, which assesses the alignment of the model's suggested repairs with the ground truth based on an exact match criterion.

\subsubsection{Joint Evaluation}The Joint Repair Accuracy provides an integrated assessment of localization and repair effectiveness. It repairs entire programs by combining token-level localization with token-level repair, selecting the top 1 candidate based on exact match criteria. It evaluates repair accuracy across the whole program and uses the CodeBLEU [19] score to assess the semantic similarity between the correct and repaired programs.
 
\section{Experiment Results}
\subsection{RQ1. How effective is LecPrompt in localizing and repairing logical errors compared to other state-of-the-art method?} 
We commenced by employing the text-davinci-003 model to identify logical errors within the QuixBugs-LE dataset, both at the token and line levels, utilizing log probability metrics. This approach is a zero-shot method without fine-tuning. The outcomes of our localization efforts compared with state-of-the-art method are presented in Table I.

\begin{table}[h!]
    \centering
    \begin{tabularx}{1\linewidth}{cXXXX} 
        \toprule
         & \multicolumn{2}{c}{Python} & \multicolumn{2}{c}{Java} \\
        \midrule
        & Token-level & Line-level & Token-level & Line-level \\
        \midrule
        CoCoNuT & --- & 24.8\% & --- & 21.6\% \\ 
        LecPrompt & 33.1\% & 53.7\% & 35.6\% & 51.8\% \\ 
        \bottomrule
    \end{tabularx}
    \caption{Localization accuracy comparison for Python and Java of QuixBugs-LE. '---' denotes data not provided because the model's inability to localize error at the token-level.}
    \label{tab:localization_accuracy}
\end{table}

In Table II, traditional models like CoCoNuT and RoBERTa are outperformed by the specialized CodeBERT variants, especially the MLM-\textbf{FT} and MLM-\textbf{PT} versions. These two exhibit notably higher repair accuracies across top 1, 5, and 10 rankings for both Python and Java, underscoring the benefits of tailored pre-training and fine-tuning for code tasks. Their superior CodeBLEU scores further affirm the quality of their repairs. Interestingly, most models tend to perform slightly better on Python than Java, potentially reflecting Python's easier repairability or dataset nuances.

\begin{table*}[h]
    \centering
    \begin{tabular}{cccccccccccc}
        \toprule
        & \multicolumn{5}{c}{Python} & & \multicolumn{5}{c}{Java} \\
        \cline{2-6} \cline{8-12}
        & \multicolumn{3}{c}{Repair} & \multicolumn{2}{c}{Joint} & & \multicolumn{3}{c}{Repair} & \multicolumn{2}{c}{Joint} \\
        \cline{2-4} \cline{5-6} \cline{8-10} \cline{11-12}
        Model & Top 1 & Top 5 & Top 10 & Full-Repair & CodeBLEU & & Top 1 & Top 5 & Top 10 & Full-Repair & CodeBLEU \\
        \midrule
        CoCoNuT & -- & -- & -- & 22.4 & 68.85 & & -- & -- & -- & 18.7 &  64.74\\
        RoBERTa & 26.17 & 43.92 & 49.53 & 12.0 & 64.47 & & 26.86 & 40.72 & 50.22 & 9.34 & 59.36 \\
        CodeBERTa & 29.9 & 62.61 & 68.22 & 10.6 & 65.48 & & 33.72 & 51.93 & 60.06 & 11.74 & 61.3 \\
        CodeBERT-MLM & 39.25 & 55.14 & 64.48 & 15.2 & 68.93 & & 37.32 & 54.48 & 66.8 & 12.93 & 62.72 \\
        \midrule
        CodeBERT-MLM-\textbf{FT} & 76.79 & 88.14 & 94.02 & 27.6 & 78.22 & & 71.22 & 85.35 & 93.2 & 25.3 & 75.2 \\
        LecPrompt & 74.58 & 86.28 & 92.1 & 27.4 & 76.49 & & 69.23 & 83.52 & 94.66 & 24.7 & 73.68 \\
        \bottomrule
    \end{tabular}
    \caption{Model performance comparison on QuixBugs-LE for Python and Java. '---' indicates unavailable data due to the model's inability to repair at the token level. Repair accuracy covers the top 1, 5, and 10 candidate accuracies (\%). Joint accuracy covers full-repair accuracy (\%) and the CodeBLEU score.}
    \label{tab:model_performance}
\end{table*}

In Table III, we observe varied accuracies across error types. For instance, "Operator Misuse" boasts a high full-repair rate of 93.2\%, while "Wrong Literal Value" is at a modest 66.4\%. Interestingly, the former type shows a significant gap between locating the error and actually repairing it, suggesting ease in detection but high proficiency in correction. In contrast, errors like "Variable Misuse" display comparable accuracies in both aspects, denoting consistent performance. However, certain errors, such as "Wrong Literal Value," present challenges in both detection and repair, hinting at either their intrinsic complexity or limited representation in the dataset.

\begin{table}[h]
    \centering
    \begin{tabular}{ccc}
        \toprule
        \textbf{Error Type} & \multicolumn{2}{c}{\textbf{LecPrompt}} \\
        \cmidrule(lr){2-3}
        & \textbf{Token-level Loc} & \textbf{Repair (Top 1)} \\
        \midrule
        Operator Misuse & 35.2 & 93.2 \\
        Variable Misuse & 33.4 & 70.6 \\
        Wrong Literal Value & 20.2 & 66.4 \\
        Broken Control Flow & 27.4 & 78.1 \\
        Incomplete Statements & 27.7 & 70.2 \\
        Redundant Operations & 21.6 & 69.2 \\
        \bottomrule
    \end{tabular}
\caption{LecPrompt's token-level localization and token repair accuracy with top-1 candidate (\%) across various logical error types in QuixBugs-LE.}
\label{tab:performance}
\end{table}

\subsection{RQ2. How does the design of the soft prompt and the mask percentage affect LecPrompt's performance?}

\textbf{Soft prompt initialization.} When it comes to soft prompt initialization, there are several different methods that can be used. (1) One approach is to initialize the prompt embeddings by copying the frozen CodeBERT embeddings. This approach is known as CodeBERT-init. (2) Another approach is to randomly initialize the prompt embeddings, which is known as Random-init. (3) A third approach is to assign all the prompt embeddings to a single value, typically 1. This approach is known as Const-init. (4) Lastly, a approach is to initialize the prompt embeddings with the same tokens, typically a sequence of special tokens that indicate the task and model parameters, and this approach is known as Token-init. Table IV shows the performance of different soft prompt initialization methods in terms of top-1 token prediction accuracy of CodeBERT-MLM-\textbf{PT}. The four methods compared are CodeBERT-init, Random-init, Const-init, and Token-init. The results indicate that initializing the prompt embeddings with the same tokens (Token-init) achieves the highest accuracy at 46.42\%.

\begin{table}[h]
    \centering
    \begin{tabular}{cc}
        \toprule
        Initialization Method & Accuracy (\%) \\
        \midrule
        CodeBERT-init & 44.58 \\
        Random-init & 42.13 \\
        Const-init & 43.82 \\
        Token-init & 46.42 \\
        \bottomrule
    \end{tabular}
    \caption{Soft prompt initialization}
    \label{tab:initialization_accuracy}
\end{table}

\textbf{Soft prompt place.} When it comes to the placement of the soft prompt, there are several different options to consider. (1) One approach is to insert the prompt at the front or back of the frozen embeddings. (2) Another approach is to place the prompt in the middle of the frozen embeddings. (3) A third option is to randomly insert the prompt at various locations within the frozen embeddings. By investigating the performance of models with soft prompts placed in different locations, we can gain insights into which placement method is most effective for a given task or dataset. The results are shown in Table V. There seems to be no significant difference in accuracy between the different positions of the soft prompt, except for the prompt in the back which yields slightly higher accuracy. However, the prompt randomly inserted into frozen embeddings significantly decreases the accuracy. These results suggest that the position of the soft prompt may not be crucial, but its consistent presence is important for accurate predictions.

\begin{table}[h]
    \centering
    \begin{tabular}{cc}
        \toprule
        Prompt Position & Accuracy (\%) \\
        \midrule
        Prompt in the front & 44.58 \\
        Prompt in the back & 44.81 \\
        Prompt in the mid & 44.62 \\
        Prompt random insert & 38 \\
        \bottomrule
    \end{tabular}
    \caption{Soft prompt position}
    \label{tab:prompt_position_accuracy}
\end{table}

\textbf{Soft prompt length.} We carried out a prompt-tuning experiment where we varied the prompt length in increments of 10, ranging from 0 to 100. As shown in Figure 6 (a), the results indicate that a prompt length of 40 tokens yields the best performance.


\textbf{Masking percentage.} Masked language models conventionally use a masking percentage of 15\% due to the belief that less masking would guarantee sufficient context to learn good representations, and more masking would make training too expensive. We experimented with different mask percentages in 5\%, 15\%, 25\%, \ldots, 95\% on CodeBERT-MLM-\textbf{PT}. As shown in Figure 6 (b), we find that masking 55\% of input tokens can outperform the 15\% percentage, as measured by fine-tuning on the downstream top-1 token prediction task.


\begin{figure}[!t]
    \centering
    \subfloat[]{ 
        \includegraphics[width=0.45\columnwidth]{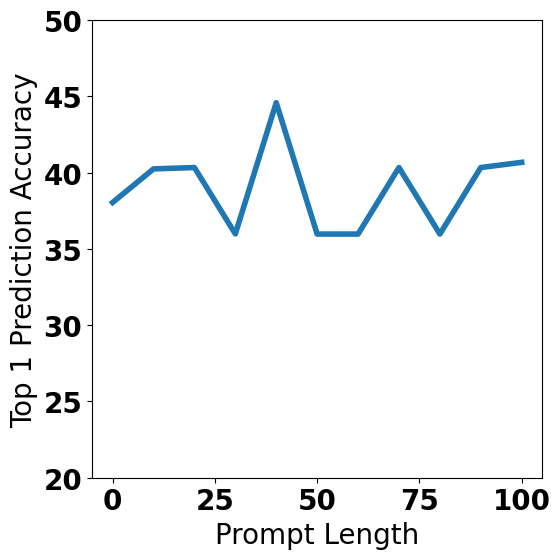}
        \label{fig:p1a}
    }
    \hfil
    \subfloat[]{ 
        \includegraphics[width=0.45\columnwidth]{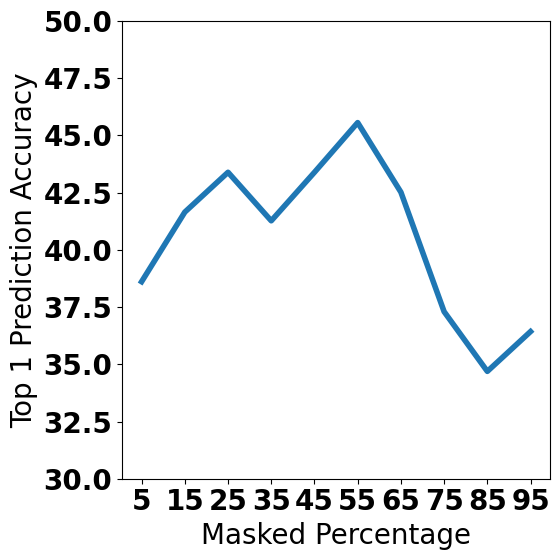}
        \label{fig:p1b}
    }
    \caption{Top-1 prediction accuracies for different parameters. (a) Top-1 prediction accuracy of CodeBERT-MLM-\textbf{PT} vs. soft-prompt length from 0 to 100. (b) Top-1 token prediction accuracy of CodeBERT-MLM-\textbf{FT} vs. masked percentage.}
    \label{fig:p1}
\end{figure}

\subsection{RQ3. What advantages does the soft prompt method offer, and what are its limitations?}

\textbf{Space-efficient.} Soft prompts offer significant advantages in terms of space efficiency. We investigate the space efficiency of prompt fine-tuning compared to conventional fine-tuning methods. With a size of 128.88 KB, the soft-prompt parameters are considerably smaller than the 316 M parameters present in the original CodeBERT. The storage space for CodeBERT-MLM-\textbf{PT} increases by only 128.88 KB for each checkpoint. It is almost difficult to observe the rise in space spend from the figure. This is because the CodeBERT parameters are frozen, and we only need to preserve the soft-prompt parameters. Initially, we store the frozen CodeBERT parameters for the first time, and during subsequent soft-prompt fine-tuning steps, we only focus on preserving the soft-prompt parameters. It is worth noting that while the size of the CodeBERT parameters is 316M, its checkpoint size is 1.5 GB because it also includes files such as vocab, optimizer, and scheduler.


\textbf{Few-shot Scenario.} We investigate the performance of soft-tuning on few-shot scenario. Figure 7 shows the performance of repair accuracy with top 1 candidate on Codeparrot-filter subset. We selected subsets of 200, 400, 600, 800, and 1000 from QuixBugs-LE for evaluation. In each case, prompt-tuning consistently outperformed the fine-tuning method, highlighting the advantages of the soft prompt-tuning approach in few-shot scenarios.

\begin{figure}[!t]
    \centering
    \subfloat[Python]{ 
        \includegraphics[width=0.47\columnwidth]{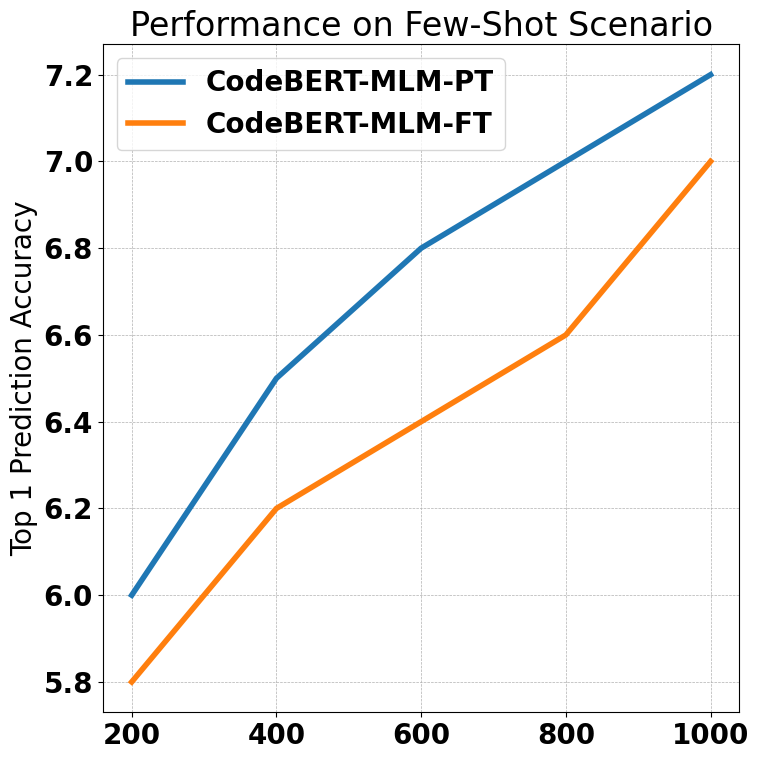}
        \label{fig:python}
    }
    \hfil
    \subfloat[Java]{ 
        \includegraphics[width=0.47\columnwidth]{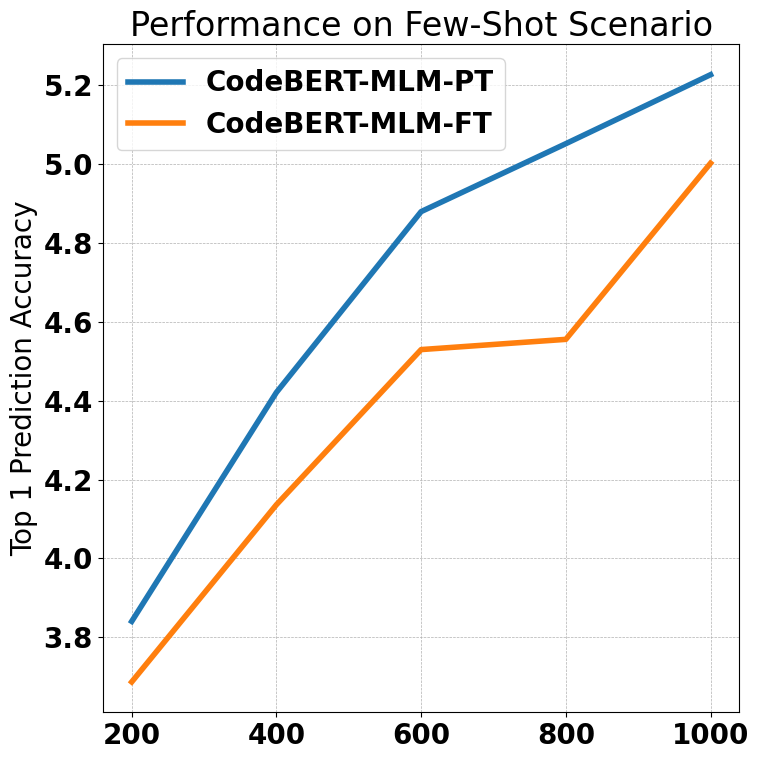}
        \label{fig:java}
    }
    \caption{Result of fine-tuning and soft prompt-tuning on top 1 candidate prediction accuracy on few-shot scenario.}
    \label{p2}
\end{figure}

\textbf{Fine-tune steps.} Fine-tune steps are also another reason to affect the performance of the soft prompt method, as we can see. As shown in Figure 8, soft prompt fine-tune needs more adaption steps to catch up with the performance of CodeBERT-MLM-\textbf{FT}.

\begin{figure}[htbp]
\centerline{\includegraphics[scale=0.4]{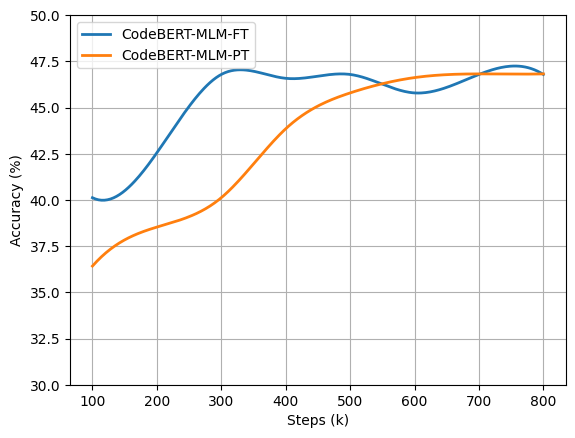}}
\caption{Top-1 token prediction accuracy of CodeBERT-MLM-\textbf{FT} and CodeBERT-MLM-\textbf{PT} vs. training steps for fine-tune and soft prompt-tuning. Soft prompt-tuning requires more training time.}
\label{p1}
\end{figure}

\subsection{RQ4. What implications arise if the localization process misidentifies the location of a logical error?}
The practical application of any code repair tool hinges not just on its ability to correctly repair code but also its robustness when faced with imperfect information. Considering that in real-world scenarios, error localization may not always be accurate, we evaluated the resilience of CodeBERT in handling such inaccuracies. In Table VI, we demonstrate CodeBERT's capability to make accurate repair predictions even if the token-level localization is off the mark. More specifically, the evaluation simulates a situation where CodeBERT is tasked with predicting hidden tokens when the masked tokens don't align with the real error locations. For this, we masked a random 15\% of tokens, deliberately excluding the genuine error positions, and probed the model to predict these hidden tokens.

Over 70\% of programs from the QuixBugs-LE python dataset managed to achieve a top-1 prediction accuracy above 90\% post this random masking. This heightened accuracy, especially when juxtaposed against fixing logical error tokens, can be attributed to the nature of the masking process itself. The random masking often includes tokens that are inherently easier to predict. In essence, if we were to attribute a 'difficulty coefficient' to each token, the ones containing genuine logical errors would undoubtedly be harder to rectify, whereas the inadvertently masked tokens tend to be more predictable.This experiment underscores a pivotal observation: even with partially misguided token indications, there's a reasonable probability of restoring the token correctly. It paints a picture of a model that doesn't solely rely on localization hints, but rather, possesses a deeper understanding of code semantics.

\begin{table}[h!]
    \centering
    \begin{tabular}{cc}
        \toprule
        Model & Accuracy (\%) \\
        \midrule
        CodeBERT-MLM-\textbf{FT} &88.37 \\
        CodeBERT-MLM-\textbf{PT} &86.56 \\
        \bottomrule
    \end{tabular}
    \caption{Random masks}
    \label{tab:model_accuracy}
\end{table}




\section{Discussion and Future Work}

Through our studies, we've identified several key areas that could pose threats to the validity of our current approach. These will serve as focal points for our future direction.

\subsection{Limited Dataset}
The dataset we utilized was iteratively generated through the fill-mask task of CodeBERT. The masked language model (MLM) pre-training of CodeBERT restricts it to operate on a single token at a time. Consequently, our method only attempts to introduce errors into one token at a time. However, semantic changes caused by successive multiple tokens may be closer to real-world logical errors. Such error types  are challenging for CodeBERT to produce. As a future direction, we plan to employ CodeT5 [13] instead of CodeBERT. Unlike CodeBERT, CodeT5 facilitates setting a mask span, enabling it to mask multiple tokens simultaneously. This would lead to a more realistic spectrum of logical errors.

\subsection{Limited Localization}
We currently employ a zero-shot approach using OpenAI's text-davinci-003 to compute log probabilities and perplexities. While this model performs exceptionally on mainstream languages like Python, Java, and JavaScript, it may not be as competitive for low-resource languages, such as Solidity. To address this, there is a need to leverage open-source large language models (LLMs) and fine-tune them for such low-resource codebases. Future works will explore fine-tuning on models like GPT-J [43], GPT-Neo [44], PolyCoder [45], and GPT-NeoX [46] to achieve a more adaptable localization mechanism.

\subsection{Limited Repair}
Program repair at the token level can be categorized into three distinct operations: (1) insertion of correct token(s), (2) deletion of error token(s), and (3) substitution of error token(s) with the correct ones. While LecPrompt excels at fixing logical errors that require replacements, it falters when confronted with more intricate logical anomalies. However, the high accuracy of line prediction presents a novel avenue to explore. We aim to harness LLMs to regenerate an entire line based on predicted lines, transforming the logical error correction challenge into a line-level code completion task. Moreover, our observations from the top 1/5/10 token candidates indicate that lots of correct tokens are present within the top 5/10 candidates produced by LecPrompt. Yet, our current strategy solely relies on the top 1 candidate. Optimizing the fill-mask process to effectively harness all available candidates will be our subsequent research focus.

\section{Related Work}

\subsection{Automatic Logical Error Program Repair}

Over the years, various methodologies have emerged to tackle logical errors in programming. AVATAR [36], introduced as an automated program repair technique, employs machine learning to generate and rank candidate patches for Java programs, showcasing its efficacy on the Defects4J benchmark. Around the same time, FixML emerged [37], adeptly diagnosing and rectifying errors in functional programming assignments by melding statistical error-localization with type-directed program synthesis. Neural Attribution for Semantic Bug-Localization (NBL) then came into the spotlight, leveraging deep learning to guide students in mending their C programs [38]. Song et al. further enriched the field by devising a technique that automatically detects discrepancies between reference functional programming assignments and student submissions [39]. CoCoNuT is an automated program repair technique that uses ensemble learning and context-aware neural machine translation to fix errors in multiple programming languages [40]. It has been evaluated against state-of-the-art APR techniques and approved to be able to repair logical errors. More recently, Yoshizawa et al. brought forth an innovative iterative model, addressing the longstanding challenges compilers and IDEs face with logical errors in source code [41].

\subsection{Pre-trained Language Models for Programming Languages}

In recent years, pre-trained language models have gained significant attention in the field of natural language processing (NLP) for their impressive performance on various NLP tasks, including language translation, text summarization, and sentiment analysis. As a result, there has been growing interest in applying these models to programming languages. By training on code or documentation and fine-tuning for tasks such as code completion, bug detection, and code generation, pre-trained language models can potentially improve software development processes. Several studies have investigated the use of these models for programming languages, showing promising results and potential for program understanding and program generation. CodeBERT [12], CodeT5 [13], PolyCoder [14], Codex [15], AlphaCode [16], CodeGen [17], and PaLM-Coder [18] are some examples of pre-trained language models that have been developed for programming languages. CodeBERT is proposed by Zhangyin Feng et al. in 2020 as a bimodal pre-trained model for programming languages (PL) and natural languages (NL). It is designed to learn general-purpose representations that support NL-PL downstream applications such as natural language code search and code documentation generation [12]. CodeBERT is trained with a hybrid objective function that incorporates the task of replaced token detection, which enables the use of both bimodal NL-PL data and unimodal data. Our LecPrompt leverages CodeBERT to conduct token-level logical error correction tasks.


These pre-training tasks enable the language model to learn a general representation of language that can be fine-tuned for various downstream tasks, such as natural language processing, text classification, and text generation. By leveraging pre-training, language models can achieve state-of-the-art performance on a wide range of natural language processing tasks. Masked-language modeling involves randomly masking certain tokens in the input sequence and training the model to predict the masked tokens based on the surrounding context [12]. This task encourages the model to learn a deeper understanding of the relationships between words in a sentence, including syntax, semantics, and context.

\subsection{Prompt Tuning}
Prompt tuning in language models can be categorized into hard and soft prompts. Hard prompts, such as PET [22], modify the input text into specific templates, often in a cloze-style, guiding the model to fill in the gaps. Another instance, AutoPrompt [42], employs unique sentence structures embedded with trigger tokens to guide model predictions. On the other hand, soft prompts focus on optimizing continuous vectors for desired outputs. Prefix-Tuning [23], for example, integrates a continuous vector into each layer of a static language model, adjusting only these vectors during training. Another variant, P-Tuning [25], automates the search for the best continuous prompt vector, fine-tuning GPT models for natural language understanding tasks. In essence, hard prompts involve structured input modifications, while soft prompts leverage flexible vector adjustments to direct model outputs.

\section{Conclusion}
In this paper, we introduced LecPrompt, a prompt-based approach for automatically correcting logical errors in programming languages. LecPrompt leverages the power of pre-trained large language models, specifically CodeBERT, to provide efficient and accurate logical error localization and repair at the token level. By mapping the downstream task of logical error correction to Masked Language Modeling (MLM), LecPrompt can effectively utilize pre-learned context understanding of programming languages. The soft-prompt embeddings in LecPrompt's architecture enable the model to adapt to the specific task of correcting logical errors without fine-tuning the entire parameters. Additionally, we showed that by using a prompt-based method, we can achieve comparable performance with fine-tuning while adjusting significantly fewer parameters. Our code is available at https://github.com/Zhenyu2049/LecPrompt.






\end{document}